\documentclass[prl,twocolumn, preprintnumbers,amsmath,amssymb, superscriptaddress]{revtex4}

\usepackage{amssymb,amsfonts,amsmath}

\usepackage{amsmath}
\usepackage{amssymb}
\usepackage{times}
\usepackage{mathrsfs}
\usepackage{graphicx}
\usepackage{dcolumn}
\usepackage{bm}
\usepackage{color}
\usepackage{epstopdf}

\newcommand{\be}{\begin{equation}}
\newcommand{\ee}{\end{equation}}
\newcommand{\bea}{\begin{eqnarray}}
\newcommand{\eea}{\end{eqnarray}}

\newcommand{\Ground}{g}
\newcommand{\Excited}{c}
\newcommand{\Pg}{P_{\Ground}}
\newcommand{\Pex}{P_{\Excited}}
\newcommand{\Rgex}{\Gamma_{\Ground \to \Excited}}
\newcommand{\Rexg}{\Gamma_{\Excited \to \Ground}}

\newcommand{\kb}{k_{\rm B}}
\newcommand{\avg}[1]{{#1}}

\newcommand{\volume}{\mathcal{V}}

\newcommand{\ns}{n}
\newcommand{\nsg}{n_g}
\newcommand{\nd}{N}
\newcommand{\ndg}{N_g}
\newcommand{\HM}{H}
\newcommand{\ECs}{E_{\rm s}}
\newcommand{\ECd}{E_{\rm d}}
\newcommand{\ECo}{J}
\newcommand{\ECom}{J_-}
\newcommand{\ECop}{J_+}

\newcommand{\Rs}{R_{\rm s}}
\newcommand{\Rd}{R_{\rm d}}

\newcommand{\TL}{T_{\rm L}}
\newcommand{\TR}{T_{\rm R}}
\newcommand{\Ts}{T_{\rm s}}
\newcommand{\TSO}{T_{0, \rm s}}
\newcommand{\Tdet}{T_{\rm d}}
\newcommand{\TdetO}{T_{0, \rm d}}
\newcommand{\TmO}{T_{0, m}}

\newcommand{\varpm}{\mathbin{\vcenter{\hbox{\oalign{\hfil$\scriptstyle-$\hfil\cr\noalign{\kern-.3ex} $\scriptscriptstyle({+})$\cr}}}}}

\newcommand{\Rm}{\Gamma_{m\rm, I}}
\newcommand{\TD}{T_{\rm D}}
\newcommand{\RD}{\Gamma_{\rm d}}
\newcommand{\DI}{D}
\newcommand{\TI}{T_{\rm I}}
\newcommand{\RS}{\Gamma_{\rm s}}
\newcommand{\RL}{\Gamma_{\rm L, I}}

\newcommand{\dq}{\dot q}
\newcommand{\dqLL}{\dq_{\rm L, I}^{\rm L}}
\newcommand{\dqLI}{\dq_{\rm L, I}^{\rm I}}

\newcommand{\dqmm}{\dq_{m\rm, I}^{m}}
\newcommand{\dqmI}{\dq_{m\rm, I}^{\rm I}}
\newcommand{\dqRR}{\dq_{\rm R, I}^{\rm R}}
\newcommand{\dqRI}{\dq_{\rm R, I}^{\rm I}}

\newcommand{\dqs}{\dq_{\rm s}}
\newcommand{\dqd}{\dq_{\rm d}}

\newcommand{\bs}{\beta_{\rm s}}
\newcommand{\Gammar}{g}
\newcommand{\Gammad}{D}
\newcommand{\hoR}{\Gamma_{{\rm br}}(\De)}
\newcommand{\sR}[2]{\Gamma_{#2}(#1)}

\newcommand{\hoQ}{{\dq}_{\rm br}}

\newcommand{\hoQt}{\dq_{\rm tot, br}(\De)}
\newcommand{\hoQlk}{{\dq}_{k, {\rm virt}}(\De)}

\newcommand{\hEF}{\frac{1}{2} \Delta \bar E}

\newcommand{\ebr}{\gamma(\e{})}

\newcommand{\ctR}{\Gamma_{{\rm ct}}}

\newcommand{\De}{\Delta E}
\newcommand{\RR}{\Gamma_{\rm R, I}}
\newcommand{\Pn}[1]{P_{#1}}
\newcommand{\e}{\epsilon}

\newcommand{\EF}{\Delta \bar E_k - \Delta E - \e}

\newcommand{\varmp}{\mathbin{\vcenter{\hbox{\oalign{\hfil$\scriptstyle+$\hfil\cr\noalign{\kern-.3ex} $\scriptscriptstyle({-})$\cr}}}}}

\begin{document}

\title{On-chip Maxwell's demon as an information-powered refrigerator}

\author{J. V. Koski}
\affiliation{Low Temperature Laboratory, Department of Applied Physics, Aalto University School of Science, P.O. Box 13500, FI-00076 Aalto, Espoo, Finland}
\author{A. Kutvonen}
\affiliation{COMP Center of Excellence, Department of Applied Physics, Aalto University School of Science, P.O. Box 11000, FI-00076 Aalto, Espoo, Finland}
\author{I. M. Khaymovich}
\affiliation{Low Temperature Laboratory, Department of Applied Physics, Aalto University School of Science, P.O. Box 13500, FI-00076 Aalto, Espoo, Finland}
\affiliation{Institute for Physics of Microstructures, Russian Academy of Sciences, 603950 Nizhni Novgorod, GSP-105, Russia}
\author{T. Ala-Nissila}
\affiliation{COMP Center of Excellence, Department of Applied Physics, Aalto University School of Science, P.O. Box 11000, FI-00076 Aalto, Espoo, Finland}
\affiliation{Department of Physics, Brown University, Providence RI 02912-1843, U.S.A.}
\author{J. P. Pekola}
\affiliation{Low Temperature Laboratory, Department of Applied Physics, Aalto University School of Science, P.O. Box 13500, FI-00076 Aalto, Espoo, Finland}

\begin{abstract}
We present an experimental realization of an autonomous Maxwell's Demon, which
extracts microscopic information from a System and reduces its entropy
by applying feedback. It is based on two capacitively coupled single
electron devices, both integrated on the same electronic circuit. This
setup allows a detailed analysis of the thermodynamics of both the Demon
and the System as well as their mutual information exchange. 
The operation of the Demon is directly observed as a temperature drop in the
System. We also observe a simultaneous temperature rise in the Demon arising from
the thermodynamic cost of generating the mutual information.
\end{abstract}


\maketitle

Thermodynamic processes are governed by fundamental laws, of which the first, conservation of energy, is paramount in all fields of physics and cannot be violated at any level of description known to date. The Second Law in turn states that entropy, the measure of disorder, of a closed system cannot decrease. This has most important consequences, such that heat flows from hot to cold, irreversible processes must dissipate work, and devices of perpetual motion are impossible. To challenge this law, James Clerk Maxwell presented a thought experiment in 1867 of a ``finite being'' capable of accurately measuring the velocity of molecules \cite{MD}. It would act between two separated reservoirs, permitting only fast molecules to enter one reservoir, while allowing only the slow ones to the other. Under such a process heat is transferred from cold to hot, apparently in violation of the Second Law. This idea, coined as "Maxwell's demon" by Lord Kelvin, has over a century spurred further research on the relation between information and energy establishing quantitative relations \cite{Szilard29,Landauer61,Landauer88,Sagawa2008,Sagawa2010,Mandal2012,Esposito2012,Deffner2013, Strasberg2014,Barato2014,Ciliberto2015}. Ongoing progress in nanotechnology has also provided concrete means to test such relations experimentally \cite{Serreli2006,Price2008,Thorn2008,Raizen2009,Toyabe10,Berut12,Roldan14,SE1,SE2,Parrondo2015,Chida2015}, thus re-igniting acute interest in actually constructing a Demon.

\begin{figure}
\includegraphics[width=\columnwidth]{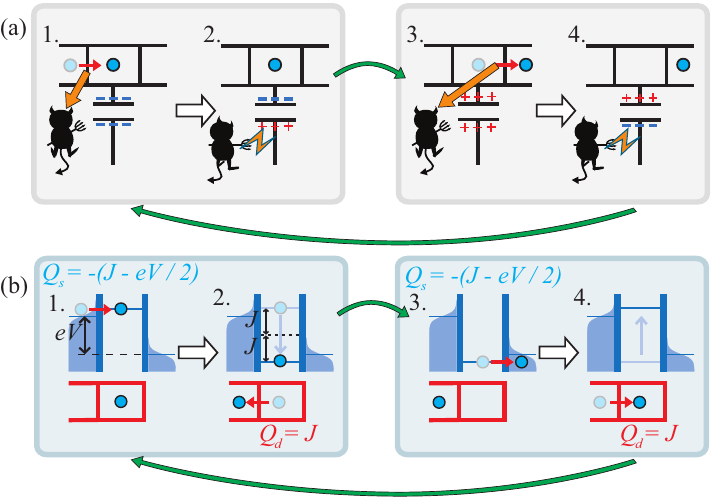}
\caption{Operation principle. (a) The Demon monitors the System (a single-electron transistor) for electrons that tunnel into (panel 1.) or out of (panel 3.) the island. It then immediately performs a feedback by applying a positive charge to trap (panel 2.), or negative charge to repel (panel 4.) the electrons. Coulomb blockade ensures that only either one or zero electrons reside in the System island. The electrons are always tunneling against the potential induced by the Demon, and therefore the System cools down. (b) Energetics of the System under voltage bias $V$ in the experimental and autonomous realization of the cycle in (a) with another single electron structure operating as the Demon. The conduction electrons of the System follow Fermi distribution, providing electrons that can overcome the energy cost $J - eV/2$, where $J$ is the coupling energy between the System and the Demon, however in doing so the System cools down by an equal amount. The energy $J$ is dissipated by the Demon as it reacts, changing the projected energy cost experienced by the electron tunneling in the System from $-J - eV/2$ to $J - eV/2$. Note that here the System island is drawn without Fermi distribution for simplicity. Also, the described operation could in princple be performed non-autonomously by externally measuring the System state, and changing the energy of the System as feedback, see e.g. \cite{Strasberg2013}.}
\label{fig_1}
\end{figure}

Recently, several theoretical proposals on configurations including both the System as well as the Demon have been presented \cite{Mandal2012, Barato2010,Strasberg2013,Shirashi2015}. Such a configuration is known as an autonomous Maxwell's demon, for the fact that the measurement and feedback operation takes place internally. Here, we experimentally realize an all-in-one Maxwell's demon, whose operation principle is cartooned in Fig.~\ref{fig_1}~(a).
The System is a single electron transistor (SET) \cite{Averin1986}, formed by a small normal metallic island connected to two normal metallic leads by tunnel junctions. The two junctions permit electron transport by tunneling, and are assumed to be identical (both with the same resistance $\Rs$). 
The Demon measures the number of electrons on the System island, and applies feedback as depicted in Fig.~\ref{fig_1}~(a). When an electron tunnels to the island, the Demon traps it with a positive charge (panels 1 and 2). Conversely, when an electron leaves the island, the Demon applies a negative charge to repel further electrons that would enter the island  (panels 3 and 4).
The System electrodes contain a reservoir of conduction electrons whose thermal excitations provide sufficiently high energy carriers to overcome the the trapping or repulsion induced by the Demon, contributing heat $Q = -\Delta E$ where $\Delta E$ is the energy cost of the tunneling event. In doing so, the System entropy decreases as $\Delta S_s = Q / \Ts$, where $\Ts$ is the System reservoir temperature, i.e. the Demon extracts information of tunneling electrons to apply feedback that causes the entropy of the System to decrease.
While the configuration resembles theoretical proposals on quantum dots \cite{Sanchez2012,Strasberg2013,Zhang2015}, and shares features with the Coulomb drag effect \cite{Solomon1989,Averin1991}, it constitutes a genuine autonomous Maxwell's Demon where only information, not heat, is directly exchanged between the System and the Demon.

Our experimental, autonomous realization of the cycle in Fig.~\ref{fig_1}~(a) relies on coupling the System island capacitively to a single electron box (SEB), a small normal metallic island connected by a tunnel junction with resistance $\Rd$ to a single normal metallic lead. Here, the SEB undertakes the role of the Demon. 
The resulting Hamiltonian is
\be \label{eq:Hamiltonian} \HM(\ns,\nd) = \ECs(\ns - \nsg)^2 + \ECd (\nd - \ndg)^2 + 2\ECo (\ns - \nsg) (\nd - \ndg), \ee
where the dynamic variables $\ns$ and $\nd$ are the net number of electrons that have entered the System and the Demon islands, respectively.
$\ECs$ and $\ECd$ are their charging energies, while $\ECo > 0$ describes their mutual Coulomb interaction and is essential for the device operation. 
The state $(\ns, \nd)$ evolves when an electron tunnels through a junction. 
$\ns$ changes to $\ns \varpm 1$ when an electron tunnels from (to) the System island.
Correspondingly, $\nd$ changes to $\nd \varpm 1$ when it transfers from (to) the Demon island.
Constant external control parameters $\nsg$ and $\ndg$ govern the System current and the coupling of the Demon to the System, respectively.
The System is voltage $V$ biased, such that the electron (with elementary charge $-e$) tunneling in the direction of (against) the voltage bias experiences an energy cost $\Delta E = \Delta \HM \varpm eV / 2$, where $\Delta \HM=\HM(\ns\pm1,\nd)-\HM(\ns,\nd)$ for changing $\ns$ is given by Eq.~\eqref{eq:Hamiltonian}. 
Similarly, for the electron tunneling in the Demon $\Delta E = \HM(\ns,\nd\pm1)-\HM(\ns,\nd)$.

The interaction between the System and the Demon is maximized by setting $\nsg = \ndg = 0.5$, producing the Hamiltonian $H(\ns,\nd)=\ECo (2\ns - 1) (2\nd - 1)/2$ and energetics depicted in Fig.~\ref{fig_1}~(b). We furthermore require $eV, k_B T \ll \ECs, \ECd$, such that only the lowest energy states of Eq.~\eqref{eq:Hamiltonian} are available, such that both $\ns$ and $\nd$ are practically limited to two possible values, $0$ and $1$. States $(\ns = 0, \nd = 1)$ and $(\ns = 1, \nd = 0)$ are charge neutral, both with energy $-J / 2$. Here, we refer to either of the states as 'ground' or $\Ground$. The state $(\ns = 0, \nd = 0)$ has an overall positive charge and $(\ns = 1, \nd = 1)$ an overall negative charge. We refer to them as 'charged' or $\Excited$, both with energy $J / 2$. Any single tunneling event will take $\Ground$ to $\Excited$ or $\Excited$ to $\Ground$, with respective $\Delta H_{\Ground \to \Excited} = J = -\Delta H_{\Excited \to \Ground}$. We assume that the System is at uniform temperature $\Ts$ while the temperature of the Demon is $\Tdet$, such that the occupation probability distribution $P_ {\ns, \nd}$ obeys $P_{0, 1} = P_{1, 0} \equiv \Pg / 2$ and $P_{0, 0} = P_{1, 1} \equiv \Pex / 2$ with $\Pg = \Rexg / (\Rgex + \Rexg)$ and $\Pex = \Rgex / (\Rgex + \Rexg)$.
Here, with notation $\ECo_\pm \equiv J \pm eV / 2$, the term $\Rexg = \Gamma_s(-\ECop) + \Gamma_s(-\ECom) + \Gamma_d(-\ECo)$ is the overall transition rate from $\Excited$ to $\Ground$, while $\Rgex = \Gamma_s(\ECom) + \Gamma_s(\ECop) + \Gamma_d(\ECo)$ is the corresponding overall transition rate from $\Ground$ to $\Excited$ as a sum of rates in the System in the direction of bias, against the bias, and the transition rate in the Demon, respectively. The transition rates are
\be \Gamma_{s / d}(\Delta E) = \frac{1}{e^2R_{s / d}} \frac{\Delta E}{e^{\Delta E / \kb T_{s / d}} - 1}. \ee
The charge current in the System is $\avg{I} = (e/2) (\Gamma_s(\ECom) - \Gamma_s(\ECop)) \Pg + (e/2)(\Gamma_s(-\ECop) - \Gamma_s(-\ECom)) \Pex$ and the total heat generation rate there is
\be \begin{split} \label{eq:System heat} \avg{\dot Q_s} = &-(\ECom\Gamma_s\left(\ECom\right) + \ECop\Gamma_s\left(\ECop\right))\Pg \\
&+ (\ECom\Gamma_s\left(-\ECom\right) + \ECop\Gamma_s\left(-\ECop\right))\Pex, \end{split} \ee
reflecting the fact that if the Demon successfully maintains a high $\Pg$ by feedback as in Fig.~\ref{fig_1}~(b), $\avg{\dot Q_s}$ is negative.
Similarly, the rate of heat generation in the Demon is
\be \label{eq:Demon heat} \avg{\dot Q_d} = -\ECo\Gamma_d\left(\ECo\right)\Pg+ \ECo\Gamma_d\left(-\ECo\right)\Pex, \ee
which in turn is positive as the Demon applies feedback on states $\Excited$ as in Fig.~\ref{fig_1}~(b).
Consider $\Ts = \Tdet \equiv T$. 
It can be shown that when $\kb T \tanh\left(\ECo / 2\kb T\right) < (\ECo / 4)(1 + \Rd / \Rs)^{-1}$, Eq.~\eqref{eq:System heat} gives negative $\avg{\dot Q_s}$, i.e. cooling, within a range of $0 < |V| < |V_{\rm max}| < 2\ECo / e$ (see Supplementary material for derivation). The entropy of the System then decreases as $\avg{\dot S_s} = \avg{\dot Q_s} / T < 0$ seemingly against the Second law, however we still get $\avg{\dot S_d} = \avg{\dot  Q_d} / T \geq - \avg{\dot Q_s} / T  = \avg{\dot S_s}$ resulting from Joule's law, $\avg{\dot Q_s} + \avg{\dot Q_d} = \avg{I} V$.

\begin{figure}[h!]
\includegraphics[width=\columnwidth]{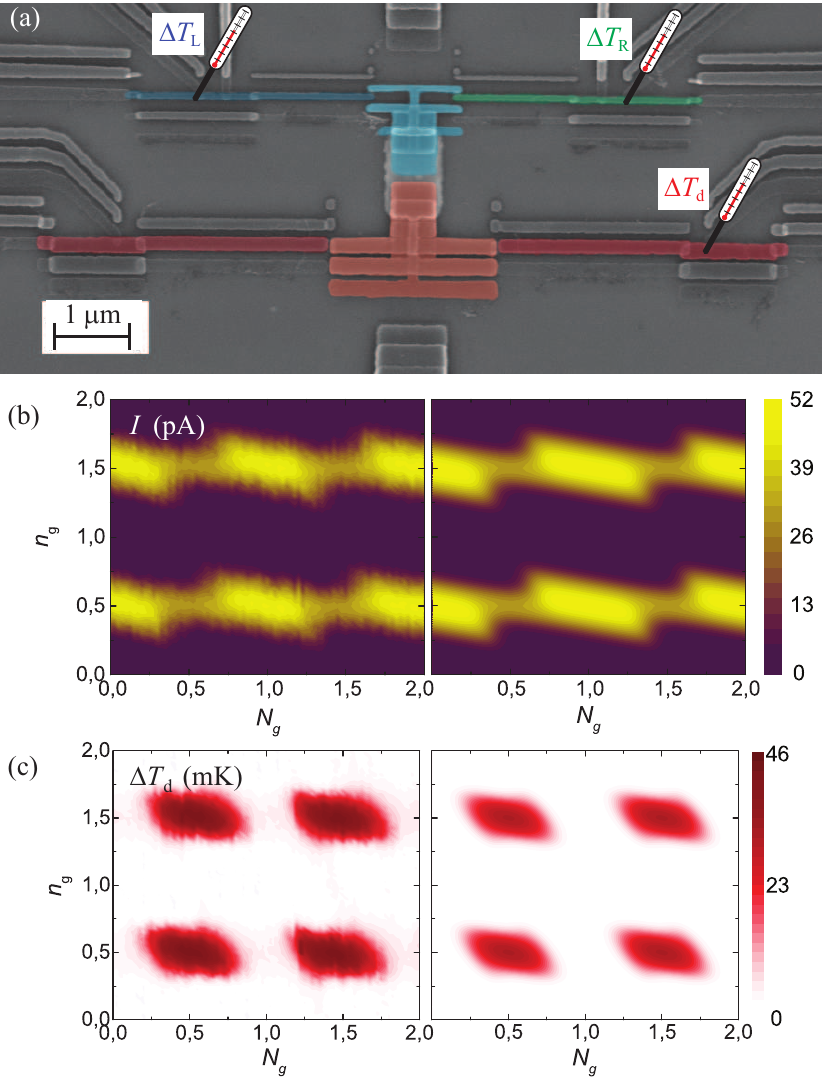}
\caption{{\bf Experimental realization.} (a) A scanning electron micrograph of the structure. False colour identifies the System island (light blue), its left lead (dark blue), and right lead (dark green), as well as the Demon island (orange) and its leads (red). The System temperature deviations from their base value, $\Delta\TL$, $\Delta\TR$, and $\Delta\Tdet$, are measured at the indicated locations (see Supplementary Material for details of measurement setup). (b) $I$ at $V = 120~\mu$V. When $\ndg$ is an integer, $I$ is modulated by $\nsg$ as in a standard SET. When $\ndg \sim 0.5$, $I$ is smaller due to Demon interaction. (c) $\Delta\Tdet$ at $V = 120~\mu$V. When $\nsg, \ndg \sim 0.5$, $\Delta\Tdet$ elevates due to the information flow between the System and the Demon.
Measured data in (b, c) are shown on the left and numerically obtained predictions on the right.}
\label{fig_2}
\end{figure}

Although energetically our device follows Joule's law, it is the information flow between the System and the Demon that permits the decrease of System entropy. The mutual information between the System and the Demon is $I_m = \ln(P_{\ns,\nd}) - \ln(P_\ns) - \ln(P_\nd)$, where $P_\ns$ and $P_\nd$ are the occupation probabilities of $\ns$ and $\nd$, respectively. As $P_{\ns = 0} = P_{\ns = 1} = P_{\nd = 0} = P_{\nd = 1} = 0.5$, mutual information changes in a tunneling event from $\Ground$ to $\Excited$ as $\Delta I_{\rm m, \Ground \to \Excited} = \ln(\Pex / \Pg)$, and for $\Excited \to \Ground$ as  $\Delta I_{\rm m, \Excited \to \Ground} = -\Delta I_{\rm m, \Ground \to \Excited}$ \cite{Sagawa2010,Abreu2012, Horowitz2014}. Tunneling events in the Demon change mutual information at the rate
\be  \label{eq:Demon information} \avg{\dot I_{\rm m, d}} = \ln \left(\frac{\Pex}{\Pg}\right) \Gamma_d\left(\ECo\right)\Pg + \ln \left(\frac{\Pg}{\Pex}\right) \Gamma_d\left(-\ECo\right)\Pex. \ee
Majority of the tunneling events in the Demon are $\Excited \to \Ground$ transitions, and since $\Pg > \Pex$, $\avg{\dot I_{\rm m, d}}$ is positive. The rate of mutual information change by the System tunneling events is $\avg{\dot I_{\rm m, s}}= -\avg{\dot I_{\rm m, d}}$. As discussed in Ref. \cite{Horowitz2014}, the System heat generation satisfies $\avg{\dot Q_s} \geq - \kb \Ts \avg{\dot I_{\rm m,d}}$ implying that the maximum amount of cooling is bound by the amount of mutual information generated by the Demon. Correspondingly, generating mutual information has a thermodynamic cost for the Demon as $\avg{\dot Q_d} \geq \kb \Tdet \avg{\dot I_{\rm m, d}}$. This can also be understood in terms of the configurational entropy $S_{\rm conf} = -\ln\left(P(\ns,\nd)\right)$ as follows~\cite{Horowitz2014}: tunneling events in the Demon bring the circuit from unlikely state \Excited~to the more probable state \Ground, decreasing $S_{\rm conf}$. At least an equivalent of heat must be dissipated to satisfy the second law. On the other hand most of the tunneling events in the System bring the setup to a more improbable state \Excited, increasing configurational entropy. The Second law then allows cooling by at most the amount of configurational entropy decreased, i.e. $-\Delta S_s \leq \Delta S_{\rm conf}$.
We note that in the limit $\Rd \ll \Rs$, $P(\ns, \nd)$ follows the thermal equilibrium distribution of the Demon. Then $\ln(P_\Ground / P_\Excited) = J / \kb \Tdet$ such that $\dot I_{\rm m, d} = \dot Q_{\rm d} / \kb \Tdet$ by Eqs.~\eqref{eq:Demon heat} and \eqref{eq:Demon information}. This implies that measurement of heat generated in the Demon is also a direct measurement of information extracted by the Demon.

\begin{figure}[h!]
\includegraphics[width=\columnwidth]{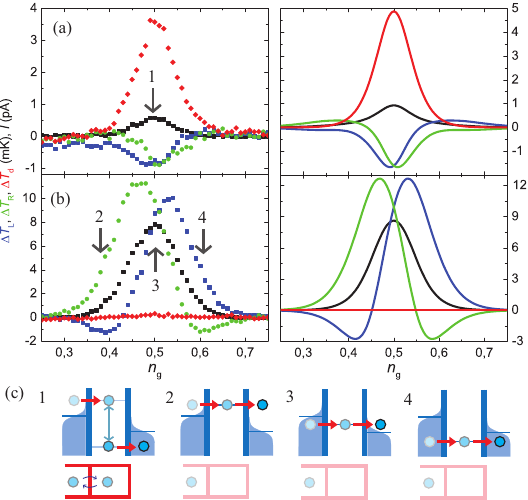}
\caption{Operation as a Maxwell's demon and as a one-sided refrigerator. Quantities shown are $I$ (black), $\Delta \TL$ (blue), $\Delta \TR$ (green), $\Delta \Tdet$ (red), with parameter values $V = 20~\mu$V, $\TSO = 77$ mK, and $\TdetO = 55$ mK. (a) Measurement at $\ndg = 0.5$ (Maxwell's demon). Both $\TL$ and $\TR$ decrease, indicating overall cooling of the System. This is justified by the mutual information transfer between the System and the Demon, which in turn generates heat in the Demon, observed as elevated $\Tdet$. 
(b) Measurement at $\ndg = 0$ (SET refrigeration \cite{Pekola2014, Feshchenko2014}). Either $\Delta\TL$ or $\Delta\TR$ can be negative, however, not simultaneously: overall heat is generated in the System. Measured data (symbols) are shown on the left and numerically obtained predictions (lines) on the right. 
(c) Energetics at different operation points, indicated as numbers in panels (a) and (b). At the operation point 1, the Demon is interacting with the System as in Fig. 1~(b). At operation points 2-4, the Demon is inactive.}
\label{fig_3}
\end{figure}

Figure \ref{fig_2}~(a) shows a scanning electron micrograph of the experimental realization of the Maxwell's demon. It was fabricated by standard electron beam lithography combined with shadow evaporation \cite{Dolan1977} of copper (normal metal) and aluminum (superconductor) metal films. Our device has the following parameters: $\ECs / \kb \simeq 1.7$ K, $\ECd / \kb \simeq 810$ mK, $ \ECo / \kb \simeq 350$ mK, $\Rs \simeq 580$ k$\Omega$, and $\Rd \simeq 43$ k$\Omega$ (two parallel junctions each with $\simeq 85$ k$\Omega$ tunneling resistance). The fully normal System and Demon junctions are realized with laterally proximized aluminum dot technique \cite{Koski11}. We determine the heat generated in the left (L) and right (R) lead of the System as well as the lead of the Demon by measuring the respective temperatures $\TL$, $\TR$, and $\Tdet$, as indicated in Fig.~ \ref{fig_2}~(a). This is achieved by reading the voltage of current-biased normal metal - insulator - superconductor junctions, see e.g. Ref. \cite{Nahum1994}. Finally, the leads of the System and the Demon are interrupted with direct contacts to superconducting leads, which permit charge transport by Andreev processes \cite{Andreev1964} but block heat transport at low temperatures. The structure is measured in a $^3$He / $^4$He dilution refrigerator at the bath temperature of 40 mK. 
Details on the device fabrication and measurement configuration are given in the Supplementary Material.

The continuous heat generation is mediated primarily by lattice phonons that couple with the conduction electron heat bath at temperature $T_{\rm L / R / d}$, contributing $\dot Q_{m, {\rm ph}}= \Sigma \volume_m (T_{0, m}^5 - T_m^5)$, $m = L$, $R$ or $d$, where $\Sigma$ is a material specific constant, $\volume_m$ is the volume of the circuit element, and $T_{0, m}$ is the base temperature~\cite{Wellstood1994}. For the left and right electrodes of the System, $\volume_{\rm L/R} \approx 2.8~\mu$m $\times$ 70 nm $\times$ 20 nm. Its island is approximately twice as large in volume. The Demon has the total volume $\volume_{\rm d} \approx 4\times3.2~\mu$m $\times$ 150 nm $\times$ 20 nm. We use $\Sigma \approx 4 \times 10^9$ Wm$^{-3}$K$^{-5}$ for Cu. 
The rate of electron tunneling ($10^6$ Hz) in our device is faster than the phonon relaxation rate ($10^4$ Hz), however it is small compared to the inelastic electron-electron relaxation rate, which is typically of the order of $10^9$ Hz \cite{Pothier1997}, allowing the electrodes to equilibriate to an effective electron temperature $T_m$ that deviates from $T_{0, m}$. Furthermore, the temperature change caused by an individual tunneling electron is sufficiently small so that $Q / T_m$ is a good approximation for the entropy change.
The temperature $T_m$ equilibrates such that the net heat generation is zero, i.e. $\avg{\dot Q_{m}} = -\dot Q_{{\rm ph}, m}$. 
The base temperature $\TmO$ is measured at $\nsg = \ndg = 0$, where the state is Coulomb blockaded to $\ns = \nd = 0$ corresponding to the energy minimum in Eq.~\eqref{eq:Hamiltonian} and no heat is generated in the circuit. Figure \ref{fig_2} (b) shows that charge current $I$ in the System modulates with $\nsg$ as in a standard SET. However, when $\ndg = 0.5$, the maximum measured current is reduced due to the feedback by the Demon. Figure \ref{fig_2}~(c) demonstrates how at $\nsg = \ndg = 0.5$, the heat generated in the Demon is maximized for extracting information of the transported electrons.

\begin{figure}
\includegraphics[width=\columnwidth]{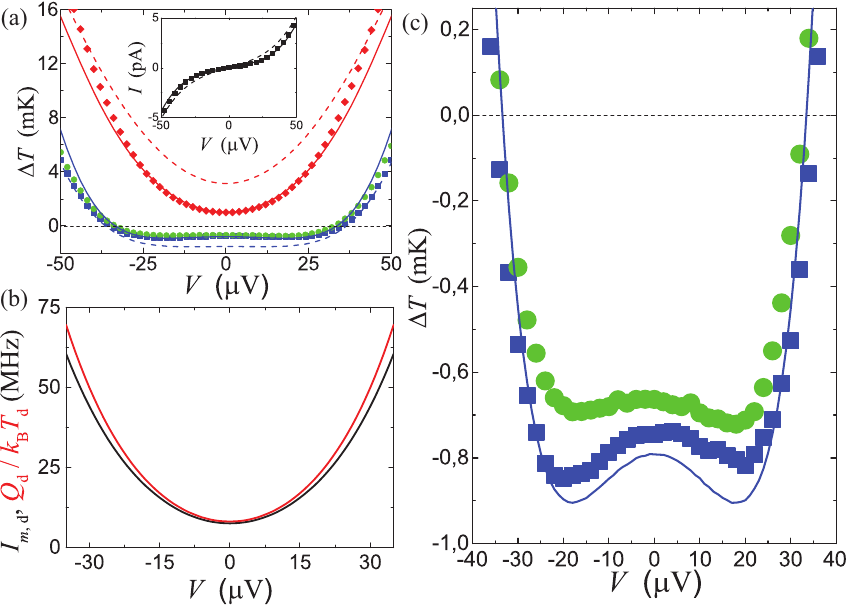}
\caption{Bias dependence. Here, $\nsg = 0.5, \ndg = 0.5$  and $\TdetO = 55$ mK. The data points (symbols) are obtained by averaging over 210 repetitions. (a) $\Delta \TL$ (blue squares), $\Delta \TR$ (green circles), and $\Delta \Tdet$ (red diamonds) with their respective prediction with $\TSO = 77$ mK (dashed lines) and $\TSO = 62$ mK (solid lines). Inset: $I$ in the same measurement. Applying voltage increases the number of electrons passing through the System and in turn the information flow between the System and the Demon. This is observed as increased $\Tdet$. (b) Numerical comparison between $\dot Q_{\rm d} / \kb \Tdet$ and $\dot I_{m, d}$, demonstrating that the two quantities match. (c) Enlarged view of the measured $\Delta \TL$ (blue squares) and $\Delta \TR$ (green circles). Increasing voltage bias further enhances the entropy decrease in the System to up to about $\pm 20~\mu$V. The model assumes a perfectly symmetric System and therefore predicts equal $\Delta \TL$  and $\Delta \TR$ with the fit $\TSO = 62$ mK (solid line). 
}
\label{fig_4}
\end{figure}

The main result of this paper is presented in Figure \ref{fig_3}~(a), showing our observation at $V = 20~\mu$V $\simeq 2J/3e$ of how the System cools down and its entropy decreases. Simultaneously, we observe how the Demon, which collects the information and immediately applies a feedback to the System, generates heat as a necessary thermodynamic cost for extracting information from the System. On the other hand, Figure \ref{fig_3}~(b) shows unchanged $\Tdet$ at $\ndg = 0$ since the Demon is effectively uncoupled from the System as its state is locked to $\nd = 0$. With that Coulomb blockade refrigeration \cite{Pekola2014,Feshchenko2014} occurs when $\nsg$ deviates from $0.5$ by causing either the left or right lead to cool down, but overall heat is generated and entropy is produced in the System.

Figure \ref{fig_4} (a) shows a measurement of current (inset) and temperatures as a function of $V$ at $\nsg = \ndg = 0.5$. Increasing voltage bias boosts electrons to pass through the System, however, at the cost of lower entropy decrease per electron. Furthermore, the risk of electrons to pass through the System without feedback control from the Demon increases, in particular via multi-electron tunneling (see Supplementary material for details). Figure \ref{fig_4} (b) compares the heat and mutual information produced by the Demon, demonstrating that they differ by less than $15 \%$ for low $V$. The data shown in Fig.~\ref{fig_4} (c) shows improvement of entropy decrease to up to $20~\mu$V, beyond which errors in the feedback process overcome the benefit of enhanced rate of electron injection. At this voltage, the cooling power on the System is estimated to be $ -\avg{\dot Q_{s}} \approx 6$ aW, while the heat dissipation in the Demon is $\avg{\dot Q_d} \approx 19$ aW. Based on the heat generation, the mutual information production rate by the Demon is then $\avg{\dot I_{\rm m, d}} \approx 25 \times 10^6$ Hz. The current is $I \approx 600$ fA, i.e. $\sim4 \times 10^6$ electrons cross the System per second. Should successful feedback be performed for every electron, the heat extracted by the Demon would be $I \times 2J / e \approx 36$ aW. Experimentally we extract $\approx$ 52$\%$, of this value i.e. this fraction of the electrons transported through the System are successfully feedback-controlled by the Demon. For efficiency at maximal cooling power, $-\dot Q_s / IV$, we the get $\approx$ 0.56.


In conclusion, we have realized and demonstrated experimentally a physically transparent autonomous Maxwell's Demon on a chip, based on coupled single-electron circuits undergoing tunneling events in a self-controlled manner. The Demon acts on the System to decrease its entropy, observed as a temperature drop. The configuration allows one to measure the effect of the Demon on the System, as well as to measure the thermodynamics of the Demon itself. The device presented here demonstrates how  information is transferred from the System to the Demon, leading to heat generation  in the Demon in amount that corresponds to the rate of information transfer. This setup constitutes a step towards autonomous information-powered nanodevices.


We thank Matthias Meschke, Felix Ritort, and Rafael S\'anchez for useful discussions. We acknowledge financial support from the Academy of Finland grants nos. 272219 and 284594, the European  Union  Seventh  Framework  Programme  INFERNOS  (FP7/2007-2013)  under  grant  agreement  no. 308850, and the V\"ais\"al\"a Foundation. We acknowledge the availability of  the  facilities  and  technical  support by Otaniemi research infrastructure for Micro and Nanotechnologies (OtaNano).


\bibliography{bibliography}

\bibliographystyle{apsrev4-1}

\clearpage

\begin{widetext}
\setcounter{figure}{0}
\makeatletter
\renewcommand{\thefigure}{S\@arabic\c@figure}

{\LARGE \bf Supplementary Material}
\newline

\section*{Fabrication and measurement setup:}

\begin{figure}[h!t]
\includegraphics[width=0.7\columnwidth]{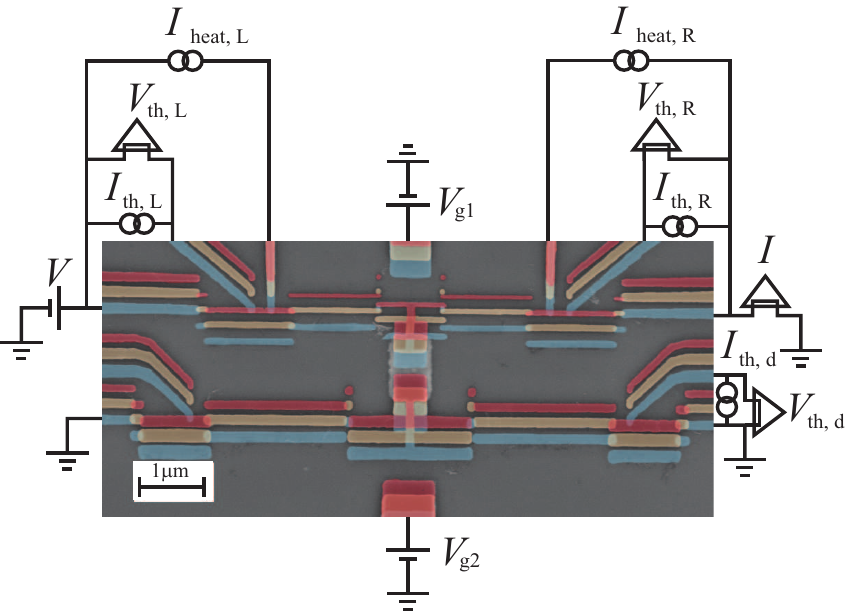}
\caption{Device fabrication and measurement configuration. False colour identifies the metal layers evaporated during the fabrication. Light blue depicts Al, and light orange is the Cu evaporated prior to oxygen exposure. Any contacts between these two layers are fully transparent. The final red is the Cu evaporated after the oxygen exposure. Any contacts to Al are tunnel junctions, whereas contacts to the first Cu layer are still highly transparent due to its slow oxidation rate. $V_{{\rm g}1}$ and $V_{{\rm g}2}$ determine the control parameters $\nsg$ and $\ndg$. By applying $I_{\rm th, L} \sim I_{\rm th, R}  \sim I_{\rm th, d}\sim 1$ pA, $V_{\rm th, L}$, $V_{\rm th, R}$, and $V_{\rm th, D}$ depend approximately linearly on the temperatures $\TL$, $\TR$, and $\Tdet$.  A $I_{\rm heat, L} \sim I_{\rm heat, R} \sim 10$ pA current is applied over the heater leads to bring $\TL$ and $\TR$ to the temperature of the System island. }
\label{S_fig_1}
\end{figure}

The System and Demon were formed simultaneously in four steps. First, a 15 nm Al (superconductor, S) layer was evaporated to form the bottommost shadow, see Fig. \ref{S_fig_1}. This was directly followed by an evaporation of 20 nm Cu (normal metal, N) layer to form the middle shadow. Any metal-to-metal contacts between Al and Cu are transparent NS contacts. After the second evaporation, the sample was exposed to oxygen, forming a thin Al$_{\rm x}$O$_{\rm y}$ layer on any Al not already covered by Cu. Finally, a second 20 nm layer of Cu was evaporated as the topmost shadow. Most of the contacts formed between Al and Cu at the third evaporation are NIS tunnel junctions with the Al$_{\rm x}$O$_{\rm y}$ as the insulator (I). However,  small Al dots with a direct contact to Cu are rendered normal due to inverse proximity effect \cite{Koski11}. This technique is used to form the NIN junctions for the System and the Demon. Copper oxidizes significantly slower than Al, thus any Cu-Cu contacts are practically transparent. The capacitive coupling between the islands of the SET and the Demon was achieved with parallel plate capacitance to an underlying Au strip evaporated prior to the SET - detector structure, covered with $\approx 30$ nm of Al$_{\rm x}$O$_{\rm y}$ achieved by atomic layer deposition (ALD). As the NIS junctions are operated as thermometers, they simultaneously induce some cooling \cite{Nahum1994} on the electrodes probed. To initialize the System in such a way that $\TL$ and $\TR$ are at the same temperature as the island, the System leads are equipped with additional current-biased tunnel junctions that induce heating to compensate for this NIS cooling effect.



\section*{Tunneling and heat generation rates}

This section gives the equations for the tunneling rates through the junctions in this device. Tunneling events may take place through the left or right System junction both with a resistance $\Rs \sim 580$ k$\Omega$, or through the left or right Demon junction both with resistance $R \sim 85$ k$\Omega$. However, as the Demon is not voltage biased, the two junctions are parallel junctions to the same potential, and can thus can be modeled as a single one with $\Rd \sim 43$ k$\Omega$. A tunneling event may take place between the left System lead (L) or right System lead (R) and the System island (I) at rate
\be \Rm(\Delta E) = \frac{1}{e^2 \Rs} \int d\epsilon f_{m}(\epsilon) (1 - f_{\rm I}(\epsilon - \Delta E)), \label{eq:LRate} \ee
where $\Delta E$ is the energy cost for the tunneling electron, $m=$~L, R, and $f_{k}(\epsilon) = (1 + \exp(\epsilon / \kb T_{k}))^{-1}$ is the Fermi-Dirac distribution of the conduction electrons in the metal $k =$~L, R, I. Similarly, the tunneling rate between the Demon lead (d) and its island (D) is
\be \Gamma_{\rm d, D}(\Delta E) = \frac{1}{e^2 \Rd} \int d\epsilon f_{\rm d}(\epsilon) (1 - f_{\rm D}(\epsilon - \Delta E)). \label{eq:fullDemonrate} \ee
We assume that $\TD = \Tdet$, in which case Eq. \eqref{eq:fullDemonrate} simplifies to
$\RD(\Delta E) \equiv \Gamma_{\rm d, \DI}(\Delta E) = (e^2 \Rd)^{-1} \Delta E/(e^{\Delta E / \kb \Tdet} - 1)$. 
Moreover, if the System is at uniform temperature implying $\TL = \TI = \TR \equiv \Ts$, the System tunneling rate simplifies to
$\RS(\Delta E) \equiv \Gamma_{\rm L, I}(\Delta E) = \Gamma_{\rm R, I}(\Delta E) = (e^2 \Rs)^{-1} {\Delta E}/(e^{\Delta E / \kb \Ts} - 1)$.

Next follow the equations for the rates of heat generation. Heat generation rate in the metal $m=$~L, R by electrons tunneling between $m$ and I is
$\dqmm(\Delta E) = (e^2 \Rs)^{-1} \int d\epsilon (-\epsilon) f_m(\epsilon) (1 - f_{\rm I}(\epsilon -\Delta E))$,
and in I it is
$\dqmI(\Delta E) = (e^2 \Rs)^{-1} \int d\epsilon (\epsilon - \Delta E) f_m(\epsilon) (1 - f_{\rm I}(\epsilon -\Delta E))$.
With $\TL = \TI = \TR = \Ts$, these simplify to
$\dqLL(\Delta E) = \dqLI(\Delta E) = \dqRR(\Delta E) = \dqRI(\Delta E) = -(2e^2\Rs)^{-1}\Delta E^2/(e^{\Delta E / \kb \Ts} - 1) \equiv \dqs(\Delta E)/2$,
where $\dqs(\Delta E)$ is the total rate of heat generation on the System, summing over the rates of the lead and the island between which the tunneling takes place. It also satisfies
\be \dqs(\Delta E) = -\Delta E \RS(\Delta E). \label{eq:HeatRateRelation} \ee
Finally, the heat generation rate by electron tunneling in the Demon (summing over d and D, with $\Tdet = \TD$) is
$\dqd(\Delta E) = -(e^2\Rd)^{-1}\Delta E^2/(e^{\Delta E / \kb \Tdet} - 1) = -\Delta E \RD(\Delta E)$.

We note that the equations above only depend on the energy cost for the tunneling process. It depends on the initial state $(\ns, \nd)$ and source and destination of the electron, as
\be \begin{split}\label{eq:EnergyCosts}
\Delta E_{\ns, \nd}^{\rm L \to I (I \to L)} &= \ECs \varmp 2\left[\ECs(\ns - \nsg) + \ECo(\nd - \ndg)\right] \varpm \frac{eV}{2} \ , \\
\Delta E_{\ns, \nd}^{\rm R \to I (I \to R)} &= \ECs \varmp 2\left[\ECs(\ns - \nsg) + \ECo(\nd - \ndg)\right] \varmp \frac{eV}{2} \ , \\
\Delta E_{\ns, \nd}^{\rm d\to D (D \to d)} &= \ECd \varmp 2\left[\ECd(\nd - \ndg) + \ECo(\ns - \nsg)\right].
\end{split}\ee

\section*{Full expressions for $I$ and $\dot Q$}
Although the states $(n, N) = (0, 0), (0, 1), (1, 0),$ and $(1, 1)$ are dominant, we also consider higher energy states in the simulations. A master equation for the configuration at steady state is written as
\begin{multline}
\frac{dP_{\ns, \nd}}{dt}=-\big[\RL(\De_{\ns, \nd}^{\rm L \to I}) + \RL(\De_{\ns, \nd}^{\rm I \to L}) + \RR(\De_{\ns, \nd}^{\rm R \to I}) 
+ \RR(\De_{\ns, \nd}^{\rm I \to R}) + \RD(\De_{\ns, \nd}^{\rm d \to D})  + \RD(\De_{\ns, \nd}^{\rm D \to d}) \big]P_{\ns, \nd} \\
+\left[\RL(\De_{\ns-1, \nd}^{\rm L \to I}) + \RR(\De_{\ns-1, \nd}^{\rm R \to I})\right]\Pn{\ns - 1, \nd} 
+ \left[\RL(\De_{\ns+1, \nd}^{\rm I \to L}) + \RR(\De_{\ns+1, \nd}^{\rm I \to R})\right]\Pn{\ns + 1, \nd} \\
+  \RD(\De_{\ns, \nd-1}^{\rm d \to D}) \Pn{\ns, \nd - 1} + \RD(\De_{\ns, \nd+1}^{\rm D \to d})\Pn{\ns, \nd + 1} = 0,
\label{eq:steadystate}
\end{multline}
which can be solved from Eq. \eqref{eq:steadystate} by noting $\sum_{\ns, \nd} P_{\ns, \nd} = 1$. In the steady state, the current through the System is the same in each cross-section and can then be written as
$I = e\sum_{n, N} \left[\RL(\De_{\ns, \nd}^{\rm L \to I}) - \RL(\De_{\ns, \nd}^{\rm I \to L})\right] \Pn{\ns, \nd}$.
The rate of heat generation in the metal $m=$ L, R is
$\dot Q_{m} = \sum_{n, N} \left[\dqmm(\De_{\ns, \nd}^{m\rm\to I}) + \dqmm(\Delta E_{\ns, \nd}^{{\rm I} \to {m}})\right] \Pn{\ns, \nd}$,
and in I it is
$\dot Q_{\rm I} = \sum_{n, N} \left[\dqLI(\De_{\ns, \nd}^{{L} \to I}) + \dqLI(\De_{\ns, \nd}^{\rm I \to {L}}) + \dqRI(\De_{\ns, \nd}^{\rm {R} \to I}) + \dqRI(\De_{\ns, \nd}^{\rm I \to {R}})\right] \Pn{\ns, \nd}$,
and finally in the Demon it is
$\dot Q_{\rm d} = \sum_{n, N} \left[\dqd(\De_{\ns, \nd}^{\rm d \to D}) + \dqd(\De_{\ns, \nd}^{\rm D \to d}) \right] \Pn{\ns, \nd}$.
The simplified forms written in Eqs.~(2, 3) in the main text are obtained in the two-by-two-state limit $n,N=0,1$ at $\nsg = \ndg = 0.5$, with uniform temperatures $\Ts$ and $\Tdet$, $\dot Q_{\rm s} = \dot Q_{\rm L} + \dot Q_{\rm I} + \dot Q_{\rm R}$, using Eq. \eqref{eq:HeatRateRelation}.

The final temperatures are determined by solving the heat balance equation for each circuit element, indexed by $m$ = L, R, I, or d, as $\dot Q_m({T_m}) + \dot Q_{m, {\rm ph}}(T_m) + \dot Q_{m, {\rm heater}}(T_m) = 0$. The term $\dot Q_{m, {\rm heater}}(T_m)$ is otherwise zero but for the metal $m=$ L, R, $\dot Q_{m\rm, heater} = \frac{L_0}{2R_{\rm h}} (T_{m}^2 - T_{0, m}^2)$ it is the heat leak through the heater junctions by Wiedemann-Franz law, $L_0 \approx 2.44 \times 10^{-8}$ W$\Omega$K$^{-2}$ is the Lorenz number,  $T_m$ is the electrode temperature, $T_{0, m}$ is the base temperature of that electrode, and $R_{\rm h} \approx 2$ M$\Omega$ is the resistance of the heater junctions.

\section*{Temperature threshold for cooling}

In this section we derive the threshold temperature for the Demon to be able to cool down the System. We consider $\nsg = \ndg = 0.5$ and uniform temperatures $\Ts$ for the System and $\Tdet$ for the Demon. In the main text Eq. (2), the rate of heat generation in the System is
\be \label{eq:System heat supp} \avg{\dot Q_{\rm s}} = -\ECom \RS\left(\ECom\right)\Pg - \ECop\RS\left(\ECop\right)\Pg
+ \ECop\RS\left(-\ECop\right)\Pex + \ECom\RS\left(-\ECom\right)\Pex,
\ee
where $J_\pm = J \pm eV / 2$.
With notation $\Gammar_x \equiv e^2 \Rs \RS(x) = x / (\exp(\bs x) - 1)$, $\Gammad_x \equiv e^2 \Rs \RD(x)$, and $\bs = 1 / \kb \Ts$, the second derivative of Eq. \eqref{eq:System heat supp} is
\begin{multline}
 \frac{d^2 \avg{\dot Q_{\rm s}}}{dV^2}\Big|_{V = 0}\times 2\Rs J(2\Gammar_\ECo + 2\Gammar_{-\ECo} + \Gammad_\ECo + \Gammad_{-\ECo}) 
=4\Gammar_\ECo \Gammar_{-\ECo}J\bs - \Gammad_{-\ECo}\Gammar_{\ECo}\left[2(1 - \bs \Gammar_{-\ECo})^2 - \ECo\bs^2 \Gammar_{-\ECo}\right] + \\ \Gammad_{\ECo}\Gammar_{-\ECo}\left[2(1 - \bs \Gammar_{\ECo})^2 - \ECo\bs^2 \Gammar_{-\ECo}\right]
 +4\bs\Gammar_\ECo\Gammar_{-\ECo}\left(-2 + \bs\Gammar_\ECo + \bs\Gammar_{-\ECo}\right)\frac{\Gammar_\ECo\Gammad_{-\ECo} - \Gammar_{-\ECo}\Gammad_{\ECo}}{2\Gammar_{-\ECo} + \Gammad_{-\ECo} + 2\Gammar_{\ECo} + \Gammad_{\ECo}}.
\label{eq:second derivative}
\end{multline}
If we further have $\Ts = \Tdet \equiv T$, we can write $\Gammad_{\ECo} = (\Rs/\Rd)\Gammar_{\ECo}$ and Eq. \eqref{eq:second derivative} reduces to
\be \frac{d^2 \avg{\dot Q_{\rm s}}}{dV^2}\Big|_{V = 0}
=\frac{\bs \Gammar_\ECo \Gammar_{-\ECo}}{2\Rs (2 + \Rs / \Rd) (\Gammar_\ECo + \Gammar_{-\ECo})}  \left[4 + \frac{\Rs}{\Rd}\left(4 - \ECo\bs\coth\left(\tfrac{1}{2}\bs \ECo\right)\right)\right].
\ee
The critical threshold to achieve cooling is $\frac{d^2  \avg{\dot Q_{\rm s} }}{dV^2}\big|_{V = 0} < 0$, i.e.
$ \bs \ECo \coth\left(\bs \ECo/2\right) > 4\left(1 + {\Rd}/{\Rs}\right), $
in which case for a finite $V$ we can achieve negative $\dot Q_{\rm s}$.
In the limit $\Rd \ll \Rs$, the numerically obtained threshold is $\kb T \lesssim 0.2611 \times \ECo$.

\section*{Multi-electron tunneling}

Tunneling beyond single-electron processes is considered in the same way as in Ref. \cite{Pekola2014}. In particular, we consider the multi-electron tunneling for the data in Fig. 4 of the manuscript, where $\nsg = \ndg = 0.5$. When $\ns = 0$ and $\nd = 1$  (or $\ns = 1$ and $\nd = 0$), any single electron tunneling event costs energy. For example, a tunneling event in the System from source to the island would bring $\ns$ from 0 to 1 with an energy cost $J - eV / 2$. However, if another electron would afterwards tunnel in the Demon, bringing $\nd$ from 1 to 0, that electron would gain $J$ in energy, resulting in a total energy gain $eV / 2$. Furthermore, in our experiment the $\Rd \simeq 43$ k$\Omega$ is close to the resistance quantum, $R_{\rm K} \approx 26$ k$\Omega$, giving rise to multi-electron tunneling in scenarios, where single electron tunneling costs energy whereas with multiple tunneling events the total energy cost would be negative. Consider an electron tunneling event with an energy cost $\De$. Right after the event, there are six possible follow-up transitions (indexed $k$), 1: $L \to I$, 2: $I \to L$, 3: $R \to I$, 4: $I \to R$, 5: d $\to$ D, or 6: D $\to$ d, each with its respective energy cost $\Delta \bar E_k$ (see Eq.~\eqref{eq:EnergyCosts}). In this context, the equations for calculating rates of tunneling and heat induced are modified by energy broadening $\ebr = \frac{\hbar}{2}\sum_k \sR{\EF}{k}$, where $\sR{\EF}{k}$ the tunneling rate of the transition $k$ (see Eqs.~(\ref{eq:LRate}, \ref{eq:fullDemonrate})). The rate of heat generation is modified as
\be \label{eq:high_orderQ} \dq_{\rm br}(\De) = \frac{1}{\pi}\int d\e \dq(\De + \e) \frac{\ebr}{\e^2 + \ebr^2},\ee
where the subscripts and superscripts are the same for $\dot q_{\rm br}$ and $\dot q$ on the left and right hand side of the equation (for example, to solve $\dot q_{\rm L, I, br}^{\rm L}$, one would insert $\dqLL$ to the right-hand side).
Similarly, the modified tunneling rates are given by
$\Gamma_{\rm br}(\De) = \pi^{-1}\int d\e \Gamma(\De + \e) {\ebr}/(\e^2 + \ebr^2)$.
The rates evaluated by the this expression are used in Eq.~\eqref{eq:steadystate} to evaluate the probability distribution $\Pn{\ns, \nd}$.
Energy conservation demands that the total rate of heat generation $\hoQt$ induced by the tunneling processes $\Gamma_{\rm br}(\De)$ satisfies $\hoQt =  -\De~\Gamma_{\rm br}(\Delta E)$. In addition to the heat generation of Eq. \eqref{eq:high_orderQ}, each junction $k$ dissipates energy $\epsilon$ as heat at a rate proportional to $\frac{\hbar}{2} \sR{\EF}{k} / (\e^2 + \ebr^2)$. Thus the additional rate of heat generated by the virtual processes for each junction (including the junction where the actual tunneling event occurs) is
$\hoQlk = \pi^{-1}\int d\e \e \Gamma(\De + \e) {\sR{\EF}{k} }/(\e^2 + \ebr^2)$,
approximately to split evenly between the electrodes shared by the junction. Now $ \hoQt \equiv  \dq^{\rm from}_{\rm br} + \dq^{\rm to}_{\rm br} + \sum_k \hoQlk = -\De~\hoR$, where 'from' and 'to' refer to 
the source and destination of the tunneling electron,
satisfies energy conservation.
Finally, we note that in the limit of $R_{p}\to \infty$ with $p = $ s, d, the effect of multi-electron tunneling should vanish: indeed with $\ebr \to 0$, $\hoQ(\De) \to \dq(\De)$ and $\hoR \to \Gamma(\De)$, whereas $\hoQlk \to 0$.

\begin{figure}[h!t]
\includegraphics[width=0.8\columnwidth]{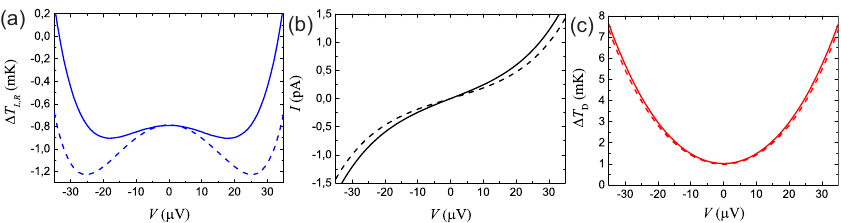}
\caption{Multi-electron tunneling. Parameters are the same as in Fig. 4 of the main manuscript. Solid lines show predictions with multi-electron tunneling, dashed ones include only single-electron tunneling.  (a) Cooling effect as a function of $V$. (b) Current through the System as a function of $V$. Multi-electron tunneling causes additional current to leak through the System, reducing the cooling power. (c) Temperature in the Demon as a function of $V$. }
\label{S_fig_2}
\end{figure}

In addition to energy broadening, we consider direct co-tunneling processes, where two electrons tunnel simultaneously. The first tunneling event takes place between pair (L,I), (R,I), or (d,D), which we indicate with A, and the second tunneling event takes place over another pair, indicated with B. The energy cost for the events are $\De_{\rm A}$ and $\De_{\rm B}$, respectively, in the case only that event took place. If both tunneling events take place, the total energy cost is $\Delta \bar E$. When $\Delta \bar E < \De_{\rm A} + \De_{\rm B}$, cotunneling events become relevant. The rate of co-tunneling is
$\ctR = \frac{\hbar}{\pi} \int d\e \frac{\Gamma_{\rm A}(\hEF + \e) \Gamma_{\rm B}(\hEF - \e)}{\left(\De_{\rm A} - \hEF - \e\right)\left(\De_{\rm B} - \hEF + \e\right)}$
and the heat dissipation rate in A is
$\dq_{{\rm A}, {\rm ct}} = \frac{\hbar}{2\pi} \int d\e \frac{\dq_{\rm A}(\hEF + \e) \Gamma_{\rm B}(\hEF - \e)}{\left(\De_{\rm A} - \hEF - \e\right)\left(\De_{\rm B} - \hEF + \e\right)}$,
where the superscript for $\dq$ is the same for left and right hand side of the equation (see the similar discussion after Eq.~\eqref{eq:high_orderQ}). The heat dissipation rate in junction B is
$\dq_{{\rm B}, {\rm ct}} = \frac{\hbar}{2\pi} \int d\e \frac{\Gamma_{\rm A}(\hEF + \e) \dq_{\rm B}(\hEF - \e)}{\left(\De_{\rm A} - \hEF - \e\right)\left(\De_{\rm B} - \hEF + \e\right)}$.
Figure \ref{S_fig_2} shows the effect of multi-electron tunneling in contrast to what would be expected if only single-electron tunneling would take place. Additional current leaks through the System without the feedback from the Demon, resulting in smaller cooling power on the System.

\section*{Gate dependence of temperatures}

Figure \ref{S_fig_4} shows how our configuration behaves as a function of $\nsg$ and $\ndg$. Figure 3 in the main text has been obtained from these type of measurements.
\begin{figure}[h!t]
\begin{centering}
\includegraphics[width=0.7\columnwidth]{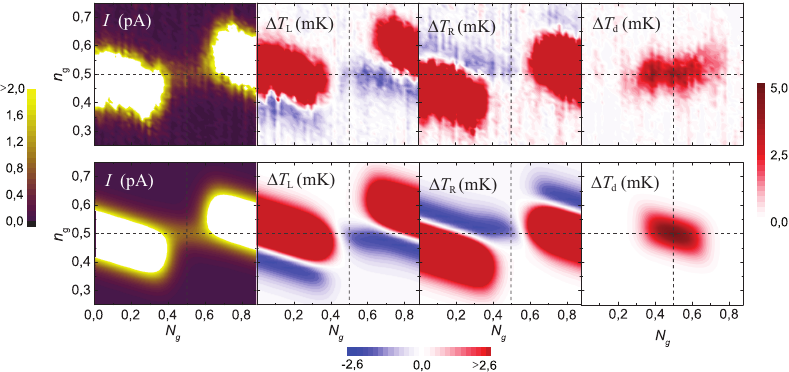}
\caption{Dependence of control parameters $\nsg$ and $\ndg$ at $\textit{V} = 20 \mu$V for current $I$ (first top and bottom panels, starting from left), $\Delta \TL$ (second panels), $\Delta \TR$ (third panels), and $\Delta \Tdet$ (last panels). Measured values are shown on top panels and simulated data are shown at the bottom. It is possible to identify the regime where the Demon begins to interact with the System (identified by the crossing point of the dashed lines), observed as suppressed maximum $I$, vanishing of heat generation in the System, and heat generation arising in the Demon.}
\label{S_fig_4}
\end{centering}
\end{figure}

\section*{Discussion on demon memory}

Many theoretical works on autonomous Maxwell's demons \cite{Mandal2012, Esposito2012, Deffner2013, Barato2014, Strasberg2014} consider systems in contact with ”information reservoirs”, represented by sequence of bits in the models. In these models the information reservoirs exchange no energy with the system.

The relation between our setup and these models can be understood by considering the two-level demon state as a memory (one bit). In the other works the sequence of bits has been initialized to a preferred initial state. That is to say, the system has been subjected to a pre-determined stream of bits. In our setup the sequence of bits is prepared by the demon itself while it operates, and not given by initialization of the stream. Because the demon has to erase the single-bit memory for the next feedback operation, there is a thermodynamic cost for this erasure, which is seen as dissipation of heat in the setup. 

As the authors in Ref. \cite{Barato2014} point out, it is possible to study the entropic interactions between the system and the information reservoir using entropic currents, which is presented by the flow of mutual information in our model. In the works \cite{Esposito2012, Strasberg2014}, the authors show in detail that in fact the entropy production and the dynamics can be understood either by considering the flow of mutual information or the model of information resource with entropy difference in the incoming and outcoming bit streams (entropy difference in the information resource).

The change in the demon state (flipping of a bit) changes the energy barriers between the system states. At this stage, the system does not yet experience transitions and therefore does not cool down, thus there is no energy transfer between the system and the demon. The carriers of heat in the setup are the electrons, and no electrons can tunnel between the system and the demon, such that direct heat exchange is in fact impossible in our setup. 

\end{widetext}

\bibliography{bibliography}

\bibliographystyle{apsrev4-1}

\end{document}